\documentclass[12pt,epsf]{article}
\usepackage{graphicx, amsmath}

\begin{document}
\sloppy
\begin{flushright}{SIT-HEP/TM-57}
\end{flushright}
\vskip 1.5 truecm
\centerline{\large{\bf Aspects of warm-flat directions}}
\vskip .75 truecm
\centerline{\bf Tomohiro Matsuda\footnote{matsuda@sit.ac.jp}}
\vskip .4 truecm
\centerline {\it Laboratory of Physics, Saitama Institute of Technology,}
\centerline {\it Fusaiji, Okabe-machi, Saitama 369-0293, 
Japan}
\vskip 1. truecm

\makeatletter
\@addtoreset{equation}{section}
\def\theequation{\thesection.\arabic{equation}}
\makeatother
\vskip 1. truecm

\begin{abstract}
\hspace*{\parindent}
Considering the mechanism of dissipative slow-roll that has been used 
in warm inflation scenario, we show that dissipation may alter usual
 cosmological scenarios associated with SUSY-flat directions.
We mainly consider SUSY-flat directions that have strong interactions
with non-flat directions and may cause strong dissipation both in
 thermal and non-thermal backgrounds. 
An example is the Affleck-Dine mechanism in which 
dissipation may create significant (both qualitative and
 quantitative) discrepancies between the conventional scenario and
 the dissipative one.
We also discuss several mechanisms of generating curvature
perturbations in which the dissipative field,
 which is distinguished from the inflaton field, can be used as the
source of cosmological perturbations. 
Considering the Morikawa-Sasaki dissipative coefficient,
the damping caused by the dissipation may be significant for many
MSSM flat directions even if the dissipation is far from thermal
 equilibrium.
\end{abstract}

\newpage
\section{Introduction}
Inflation has been considered as an important cosmological event in the
very early Universe.
In most models of inflation, a scalar field (inflaton) rolls slowly 
during inflation and the vacuum energy associated with the inflaton
potential is responsible for accelerating the expansion of the
Universe. 
Since the primeval radiation rapidly red-shifts during inflation,
the Universe during inflation would be cold.\footnote{The Hawking
temperature is intrinsic to de Sitter space. 
Non-thermal background is considered when the Hawking temperature is not
significant for the dissipation mechanism\cite{gamma-MS,
warm-inflation-original}. }   
Reheating after inflation is thus required to recover the hot Universe.
However, if there are interactions between inflaton and other fields,
the radiation background during inflation may not be completely 
red-shifted away
and the Universe may be warm during inflation. 
The warm scenario is based on the idea that 
the dissipation sources radiation and raises the 
temperature during inflation.
The scenario with the dissipating inflaton field is known 
as warm inflation\cite{warm-inflation-original}.
If the inflaton is dissipative, the equation of motion is given by
\begin{equation}
\ddot{\phi}+3H(1+r)\dot{\phi}+V(\phi,T)_\phi=0,
\end{equation}
where the subscript denotes the derivative with respect to the field
$\phi$. 
The strength of the damping is measured by the rate 
$r$ given by the dissipative coefficient $\Upsilon$ and the Hubble
parameter $H$;
\begin{equation}
r\equiv \frac{\Upsilon}{3H}.
\end{equation}
Here the effective potential of the field $\phi$ may depend on the
radiation temperature $T$, and thus the potential is expressed as
$V(\phi,T)$. 
The dissipative coefficient $\Upsilon$ of the inflaton, which is related
to the microscopic physics of the interactions, leads to the energy
conservation equation;
\begin{equation}
\dot{\rho}_R+4H \rho_R = \Upsilon \dot{\phi}^2,
\end{equation}
where $\rho_R$ is the energy density of the radiation and 
the right hand side ($ \Upsilon\dot{\phi}^2$) represents
the source of the radiation.
Scenario with $r>1$ is called 
``strongly dissipating warm inflation'' and the one with $r<1$ is called
``weakly dissipating warm inflation''.
A very confusing situation would be that inflaton is slow-rolling 
because of the strong dissipation, while the radiation redshifts away 
because the thermalization process cannot catch up with the 
rapid red-shifting.
Although inflation is not ``warm'' in this peculiar situation,
the scenario is sometimes classified into strongly dissipating warm
inflation.
According to this classification, ``warm direction'' in this paper
involves strongly dissipating field that evolves in a cold
($T\simeq0$ and $\rho_R\simeq0$) Universe. 

If we consider the dissipative field equation, the effective 
slow-roll parameters should be different from the conventional ones.  
They are given by
\begin{eqnarray}
\epsilon_w &\equiv& \frac{\epsilon}{(1+r)^2},\nonumber\\
\eta_w &\equiv& \frac{\eta}{(1+r)^2},
\end{eqnarray}
where the usual slow-roll parameters ($\epsilon$ and $\eta$) are defined
by 
\begin{eqnarray}
\epsilon&\equiv& \frac{M_p^2}{2}\left(\frac{V_\phi}{V}\right)^2,
\nonumber\\
\eta &\equiv& M_p^2\frac{V_{\phi\phi}}{V}.
\end{eqnarray}
Here the subscript denotes the derivative with respect to the dissipative
field $\phi$.

To quantify the dissipative coefficient, it would be useful to list some
past results in different models\cite{Moss:2006gt}.
The dissipative coefficient may depend on the temperature $T$ and the
expectation value of the inflaton field.
Considering interactions given by the superpotential
\begin{equation}
W=g_1\Phi X^2+g_2 XY^2,
\end{equation}
where $g_1$ and $g_2$ are coupling constants, and $\Phi, X, Y$ are
superfields whose scalar components are given by $\phi, x, y$.
Fermionic partners of the scalar components $\phi, x, y$ are $\psi_\phi,
\psi_x, \psi_y$.
During inflation, when $\phi$ is large, the field $y$ and its fermionic
partner $\psi_y$ are massless in the global supersymmetric limit.
The mediating field is $x$, which obtains large mass 
$m_x\sim g_1 \phi$ from the interaction.
In this model the dissipation is caused by the excitation of the
mediating field $x$ that decays into massless fermions.
At high temperature ($m_x \ll T$) the dissipative coefficient is
given by $\Upsilon \propto (g_1^2/g_2^2)T$, while at low temperature
($m_x \gg T$) the coefficient is given by $\Upsilon \propto
g_2^4(T^3/\phi^2)$.
In addition to the dissipation in thermal background, 
there may be significant dissipation in non-thermal background.
In fact, considering the basic mechanism of black-hole radiation in the
time-dependent background, it is straightforward to find that 
 the similar mechanism leads to the
coherent excitation of the mediating field, which appears in the
zero-temperature Universe and contributes dissipation due to the
time-dependent mass caused by $\dot{\phi}\ne 0$. 
Dissipation in the non-thermal background was studied by Morikawa and
Sasaki\cite{gamma-MS}, and later it was found that in the presence of
the mediating field the
dissipation gives $\Upsilon \propto g_1^3 g_2^2\phi$
for $T=0$\cite{gamma-MS}.
In this paper we consider $\Upsilon \simeq C_TT^3/\phi^2$ 
for the dissipation after reheating (low $T$, in the region where
 Morikawa-Sasaki coefficient is negligible) and 
$\Upsilon \propto \phi$ for the dissipation during inflation
(zero $T$, when Morikawa-Sasaki
coefficient is significant).

In past studies the dissipative slow-roll has been considered for warm
inflation.  
However, a natural expectation is that similar phenomenon (dissipative
slow-roll) may be observed  
in the evolution of other scalar fields if there are many scalar fields
in the theory. 
Based on this simple idea, we address the generic situation where many
scalar fields in the early Universe dissipate their energy during or
after inflation. 
Thermalization condition of the decay products would be a serious problem
if we discuss warm background during inflation; however it is not
important in just solving the slow-roll problem using  
 the dissipative motion in non-thermal background.
Thermalization of the decay products during inflation
(which is required for the warm background) 
would enhance the dissipative coefficient, but we modestly put
an assumption that the thermalization is not
significant during inflation.\footnote{Analytic calculations
of dissipation are usually performed in the region
$\frac{\dot{\phi}}{\phi}<H<\Gamma_{d}$,
where $\Gamma_d$ is the decay rate of the mediating field.
Dissipation beyond the above condition (e.g, preheating) may be
important, but it occurs when the slow-roll conditions
are violated\cite{preheating-papers, preheating-application}.} 
Some implications beyond this assumption will be discussed in this paper
for the enhancement of the field fluctuations for $T>H$ during
inflation. 

The most important topic in this paper is the dissipative motion of a
scalar field that is not identified with the inflaton field.
For the scalar field, we consider SUSY-flat direction.
First note that supersymmetric gauge theories appear with gauge
invariant polynomials along which the classical scalar potential in the
global supersymmetric limit vanishes.
These polynomials are related to the so-called flat directions of the
supersymmetric theory.
Flatness of the direction is not exact if supersymmetry is broken by
soft terms, or non-perturbative corrections lift the potential.
Within the minimal supersymmetric standard model (MSSM), the classical
scalar potential is given by the sum of the F and D term contributions.
Although the MSSM flat directions are usually 
parameterized by gauge invariant monomials of superfields, 
each scalar component in the monomial has charges and breaks
gauge symmetries by its vacuum expectation value (VEV). 
A generic argument of the MSSM flat directions is obviously beyond the
scope of this paper, however a simple example would be useful to gain some
insight into the dynamics related to the MSSM flat directions.
We thus consider the $H_u
L\equiv\epsilon_{\alpha\beta}H^{\alpha}_{u}L^{\beta}$  
flat direction, which is a polynomial (see the right hand side) given by
the superfield of a Higgs field ($H_u$) and the superfield of a lepton
($L$) in the MSSM action.
The F-terms are obtained from the MSSM-superpotential
\begin{equation}
W=\lambda_u QH_u \bar{u} + \lambda_d QH_d \bar{d}
+\lambda_e LH_d \bar{e}
+\mu_H H_u H_d,
\end{equation}
where the last term is the $\mu$-term.
The $\mu$-term is usually disregarded in the
calculation of the flat directions because
$\mu_H$ is of the order of the soft mass.
The flat direction associated with $H_u L$ is
\begin{equation}
H_u=\frac{1}{\sqrt{2}}\left(
\begin{array}{c}
\phi\\
0
\end{array}\right), \,\,\,\,\,\,\,
L=\frac{1}{\sqrt{2}}\left(
\begin{array}{c}
0\\
\phi
\end{array}\right),
\end{equation}
where $\phi$ is a complex scalar field that parameterizes the flat
direction.
Note that the flat direction interacts with other scalar fields, gauge
bosons and fermions.
Note also that $\phi\ne 0$ gives mass to other directions through Yukawa
couplings in the superpotential, as well as to gauge bosons.
The neutrino mass is not explicit in the MSSM action,
but it would be natural to expect that neutrinos are effectively
massless in the early Universe.
Besides neutrinos, there would be fermions 
to which mediating field (scalar fields or gauge bosons that obtain mass
from the direction $\phi\ne 0$) can decay.\footnote{
A more interesting case was discussed in ref.\cite{Matsuda:2009da},
where the gravitational decay into gravitinos is efficient for the
dissipative slow-roll in a supersymmetric hybrid inflation model.}

A very confusing property of the flat directions would be
 that they are {\bf not flat} in the early Universe.
The flat directions, which are flat in the global supersymmetric limit,
will obtain $O(H)$ mass from the supergravity corrections.
This is the famous $\eta$-problem that appears in 
supersymmetric inflationary models. 
It would be possible to forbid such corrections introducing
symmetry, but our modest assumption is that 
flat directions are not flat due to the supergravity corrections.

Any field in the early Universe has two distinctive types of evolution.
Non-dissipating evolution has been studied by many authors
for many applications.
Significant results have been obtained for the Affleck-Dine mechanism 
\cite{AD-original, AD-other, AD-iso}
and the formation of Q-balls\cite{Q-ball-original, Q-ball-other}
 using non-dissipating directions. 
Unfortunately, dissipative slow-roll has not been studied for such models.
In addition to the Affleck-Dine mechanism and Q-balls,
 dissipative motion may be important for models of generating
curvature perturbations, even if the dissipative field is not the
inflaton field.
In fact, the creation of the curvature perturbations may be due to
entropy perturbations of a light field that is distinguished from
the inflaton.
Such mechanism can be associated with fluctuations of the phase
transition\cite{IH-pt, End-Modulated, 
End-multi, End-multi-mat0} or to fluctuations in the ratio
of the energy densities\cite{curvaton-paper, matsuda_curvaton,
Topolo-curv}.   
The curvature perturbations may be generated from
fluctuations of the inflaton velocity caused by entropy
perturbations, which  in terms of the $\delta
N$ formalism\cite{Modulated-matsuda,
A-NEW} explains the creation of the curvature
perturbations at the ``bend'' of the trajectory.
The effect of warm entropy field\footnote{The adiabatic inflaton field
is defined for multi-field inflation so that it points the direction
along the inflaton motion.
Entropy fields are defined to point the directions tranverse to the
adiabatic field.} has not been studied for
the evolution of the curvature perturbations, except for a
multi-field model discussed in Ref.\cite{matsuda-warm}. 
Taking into account the dissipative motion of these scalar fields
 would alter the cosmological consequences of the scenarios.
Since cataloging these applications is not appropriate
for the purpose of this paper, we first explore the typical situation
of the Affleck-Dine mechanism using the dissipative MSSM-flat direction,
and then discuss briefly other possible applications of the warm 
scenario.

\section{Warm directions}
\subsection{Dissipative AD field for the AD baryogenesis}
Evolution of a flat direction depends on the shape of the potential.
We consider the typical form of the potential:
\begin{equation}
V(\phi)=V_I+\frac{c_s}{2}m_s^2 |\phi|^2 
+ \frac{c_H}{2} H^2 |\phi|^2 
+\left(\frac{\lambda_s m_s+\lambda_n H}{nM_p^{n-3}}\phi^n +h.c.
\right)
+|\lambda_n|^2 \frac{|\phi|^{2n-2}}{M_p^{2n-6}},
\end{equation}
where $m_s$ denotes the supersymmetry-breaking mass and 
$c$ and $\lambda$ with subscripts are dimensionless parameters.
As we are considering conventional flat directions, $n$ is an integer
greater than 4. 
The first term $V_I$ denotes the inflaton potential.
The last three terms originate from the typical supergravity action.

Let us first consider evolution of the AD field {\bf without} strong
dissipation.  
Following the conventional argument, we consider potential with
 $c_H<0$ and $H\gg m_s$ during inflation.
Then the instability near the origin leads to a global minimum at  
\begin{equation}
|\phi_0| \sim \left|\frac{c_H H M_p^{n-3}}{\lambda_n}\right|^{1/(n-2)}.
\end{equation}
Disregarding the dissipation, the field will rapidly settle down 
to a minimum at $\phi=\phi_0$.
The fluctuations of the field are negligible for $c_H\sim 1$, 
since the amplitude of the
fluctuations will quickly fall off.
The motion in the $U(1)$ direction, which is associated with rotation
of the complex scalar, is determined by the explicit
breaking term, which appears as the so-called A-term caused by
$\lambda_n H\ne0$. 
The A-term leads to a fast motion in the $U(1)$ direction, 
with a gap in the phase between the $H$-induced vacuum and the
$m_s$-induced one.
Eventually, the gap leads to rotational motion that is required for the
baryogenesis\cite{Dine:1995kz}.
Assuming that inflaton decays when $H<m_{3/2}$, where
$m_{3/2}$ is the 
gravitino mass, the baryon to entropy ratio subsequent to reheating is
given by\cite{Dine:1995kz}
\begin{equation}
\frac{n_b}{s}\simeq \frac{n_b}{n_{\phi}}\frac{T_R}{m_\phi}
\frac{\rho_{\phi}}{\rho_I},
\end{equation}
where $T_R$ and $\rho_I$ are the reheating temperature and the inflaton
mass density, $n_\phi$ and $n_b$ are AD field and baryon
number densities.
Here the energy densities are estimated at the time of
inflaton decay.
The energy density of the AD field ($\rho_{\phi}$) is calculated from
the potential $V(\phi)$.
Typical value of $n_b/n_\phi$ is $O(0.1)$, which has been obtained 
in Ref.\cite{Dine:1995kz} considering the field equation for the AD
field $\phi$ and integrating over the angle parameter.
If the AD field is non-dissipative, the ``initial'' condition of the AD
field is given by $\phi=\phi_0$, while for a dissipative
field the mechanism that determines the ``initial'' condition is 
quite different.
The difference that appears in the ``initial'' condition is the main
result of this section.

When we consider the dissipation of the field motion, we must consider
two distinctive cases for the AD-field: dissipation of the angular motion
and the radial motion.
The dissipation of the angular motion has been discussed for the
so-called ``longevity of the AD field''.
In fact, non-perturbative decay of the rotating field
has been discussed by many authors\cite{Non-pret-decat}.
When the effective ``slow-roll'' in the radial direction is simply due
to the 
rotational motion, the field starts to roll quickly as soon as
the angular velocity decreases through dissipation.
This situation is similar to a satellite losing their energy due to the
air resistance, and it is in contrast to the dissipative slow-roll
scenario. 
In contrast to the ``slow-roll'' due to the angular motion,
 the dissipative slow-roll, which is considered in this paper,
 is possible even if the rotational
motion is strongly dissipative.\footnote{Baryogenesis will be suppressed
if the angular motion is highly dissipative\cite{Non-pret-decat}.
We are not arguing this point.}

In the presence of strong ($r>1$) dissipation, the scenario of the
Affleck-Dine baryogenesis may be quite different.
The AD field may roll slowly in spite of the supergravity
corrections, and its fluctuations may become significant.
To understand the dissipative slow-roll of the Affleck-Dine field during
inflation, we follow the typical treatment of the warm inflationary
scenario\cite{warm-inflation-original, matsuda-warm}, since the
equations describing the AD field are completely the same
as the warm inflationary scenario.
We note again that the direction associated with the field motion is
flat at the tree level but obtains $O(H)$ mass due to the
supersymmetry-breaking and the supergravity corrections.
The field motion is dissipative, but the field
cannot cause warm inflation by itself.
The AD field is distinguished from the inflaton field.
We consider evolution of the AD field both during and after 
inflation.
The simplest situation is that the background radiation is negligible
($T\simeq0$ and $\rho_R\simeq0$) during 
inflation, but the radiation is not negligible after reheating.
Therefore, the dissipative coefficient may be enhanced after reheating.
We first consider the simplest situation in which dissipation during
inflation occurs in non-thermal background.

For the typical interaction of a supersymmetric flat direction we
consider the interaction Lagrangian 
\begin{equation}
{\cal L}_I =-\frac{1}{2}g^2\phi^2 \chi^2 -h\chi\bar{\psi}\psi,
\end{equation}
where $\chi$ is the scalar field (mediating field) that obtains large mass
$m_\chi \sim g\phi$ from the flat direction $\phi$, and
$\psi$ is a matter that satisfies the condition $m_\chi>2m_\psi$ for the
decay $\chi\rightarrow 2\psi$

During inflation, when $T=0$, calculation of the dissipative slow-roll
parameters is straightforward. 
Far from the origin, terms proportional to 
$\phi^n$ dominate the potential. 
Then, using $V_I\sim H^2 M_p^2$ and $m_s^2 \ll H^2$ during inflation, 
the slow-roll parameters are given by
\begin{eqnarray}
\epsilon_w &\simeq& M_p^2 
\left(\frac{\lambda_n H\phi^{n-1}}{H^2 M_p^{n-1}}\right)^2 
\frac{1}{(1+r)^2}
\sim  
\left(\frac{\lambda_n \phi^{n-1}}{H M_p^{n-2}}\right)^2 
\frac{1}{(1+r)^2}<1
\\
\eta_w &\simeq& \frac{\lambda_n (n-1)H \phi^{n-2}}{H^2 M_p^{n-3}}
\frac{1}{(1+r)^2}
\sim \frac{\lambda_n (n-1) \phi^{n-2}}{H M_p^{n-3}(1+r)^2}
<1.
\end{eqnarray}
Using the dissipative coefficient in the non-thermal background,
the dissipative rate during inflation is given by\footnote{See appendix.}
\begin{equation}
r\sim 10^{-2}N_\chi N_\psi g^3 h^2 \frac{\phi}{H},
\end{equation}
where $N_\chi$ and $N_\psi$ are the effective numbers of the mediating
field and the light fermions.
The effective slow-roll conditions give
\begin{eqnarray}
\label{slow-cond2}
\epsilon_w < 1 &\rightarrow& \phi < \phi_\epsilon \equiv
M_p\times \left(\frac{10^{-2}N_\chi N_\psi 
g^3h^2}{\lambda_n}\right)^{1/(n-2)},
\nonumber\\
\eta_w<1 &\rightarrow&\left\{ 
\begin{array}{cc}
\phi < \phi_\eta \equiv M_p
\left[\left(\frac{M_p}{H}\right)
\left(\frac{10^{-4}N_\chi^2 N_\psi^2 g^6h^4}
{\lambda_n (n-1)}\right)\right]^{1/(n-4)}
&(n>4),\\
H< M_p \times\frac{10^{-4}N_\chi^2 N_\psi^2 g^6h^4}{\lambda_n (n-1)}
 &(n=4)
\end{array}
\right. .
\end{eqnarray}
where we considered strong dissipation $r>1$.
The strong dissipation requires 
\begin{equation}
\phi > H \left(\frac{ 10^{2} }{N_\chi N_\psi g^3 h^2}\right).
\end{equation}
From the above equations, we find that the dissipative slow-roll
is common for the AD field if the initial condition is chaotic ($\phi
\sim M_p$) and the interaction coefficients ($g$ and $h$) 
are not suppressed.

Our most important result is that the initial amplitude of the AD
field is not always determined solely by the scalar potential.
For the dissipative direction, the amplitude of the flat direction 
is determined by
(1) the scalar potential, (2) the chaotic initial condition and 
(3) the interactions of the component scalar fields.
Note that only (1) is important for determining the initial condition in
the usual scenario.

After inflation and reheating, the dissipation may be enhanced by the
thermal background.
We thus consider the dissipative coefficient after reheating;
\begin{equation}
\Upsilon =C_T \frac{T^3}{\phi^2},
\end{equation}
which leads to strong dissipative region ($r>1$) given by
\begin{equation}
\phi<\sqrt{C_T M_p T},
\end{equation}
where $H\sim T^2/M_p$ is assumed.\footnote{
Our modest assumption for the dissipative motion after reheating 
is that the thermalization of the dissipative decay product
is not significant and it does not become a significant source of
radiation after reheating.
The evolution of the temperature $T$ and the radiation energy density
$\rho_R$ is thus given by the conventional
cosmological equations.}
Interestingly, the result shows that  even if the Morikawa-Sasaki
coefficient is not significant (i.e. when the ``initial'' condition
before reheating is given
by $\phi=\phi_0$), AD field that starts oscillation 
from $\phi=\phi_0$ can be trapped at $\phi\sim \sqrt{C_T M_p T}$
due to the frictional force caused by the strong dissipation.
Note that rotational oscillation of the AD field is important for the
Affleck-Dine baryogenesis, which suggests that the oscillation does not
hit the origin ($|\phi|=0$) but there is
 a minimum of $|\phi|$ for the oscillation.
Our result shows that the minimum must satisfy the condition
$|\phi|_{min}>\sqrt{C_T M_p T}$ during the rotational oscillation, which
seems very severe for the conventional scenario, in which the initial
amplitude is given by $\phi_0 \sim \left(c_H H
M_p^{n-3}/\lambda_n\right)^{1/(n-2)}$.
In fact, $|\phi_0|\gg |\phi|_{min}$ gives significant condition for the
potential with $n<6$.

If the Morikawa-Sasaki coefficient is large ($r\gg 1$)
during inflation, the field rolls slowly.\footnote{Following previous
studies of the warm inflation scenario\cite{gamma-MS,matsuda-warm-apps, 
 not-gamma-MS}, strong dissipation would be robust for
 SUSY flat directions.}
The dissipation may lead to the creation of isocurvature perturbations 
caused by the AD field.
In fact, since the AD field satisfies the dissipative slow-roll
conditions during inflation,  
 the perturbations of the AD field can exit the horizon.
To understand the significance of the perturbations, it will be
important to consider (temporarily) ``warm'' ($T>H$) background during
inflation.
Following the standard argument of warm inflation, 
the root-mean square fluctuation amplitude of the light field $\delta
\phi$ after the freeze out is obtained to be\cite{warm-inflation-original}
\begin{eqnarray}
\label{pert-phi}
\delta \phi_{\Upsilon >H}&\sim& 
(\Upsilon H)^{1/4}T^{1/2}
\sim r^{1/4}r_T^{1/2}H\,\,\,\,\,\, (\Upsilon >H),\\
\delta \phi_{\Upsilon<H}&\sim& 
(HT)^{1/2}
\sim r_T^{1/2}H\,\,\,\,\,\, (\Upsilon <H)
\end{eqnarray}
where $r_T$ denotes the ratio defined by
 $r_T\equiv T/H$.
The fluctuation is enhanced for $r_T>1$ and $r>1$.
Note that in the standard scenario of the Affleck-Dine mechanism the
 fluctuation is negligible ($\delta \phi=0$) due to the supergravity
 corrections that lead to $\eta\sim 1$ during inflation.
The discrepancy is important for the study of the isocurvature
perturbations related to the Affleck-Dine baryogenesis\cite{AD-iso,
 Topolo-curv}. 
 
\subsection{Warm directions for curvature perturbations}
In the previous section we showed that the baryon number asymmetry of
the Universe can be associated with the dissipative field.
Next we consider models of generating curvature
perturbations in which the seed of the cosmological fluctuation is
associated with the dissipative field that is distinguished from the
inflaton field.  
\subsubsection{Curvatons}
Assume that there are many components in the Universe, such as vacuum
energy, matter, oscillating field or radiation.
If the ratio of these components in the energy density is not
homogeneous in space,  
and their scaling rules are distinguishable, density
perturbations are generated during cosmological evolution.
Assume that there is an inhomogeneity 
related to the density of the field $\varphi$, and the inhomogeneity is
caused by the cosmological fluctuation of the field $\delta \varphi\ne
0$. 
Then, one can define the ratio $r_\varphi \equiv
\rho_\varphi/\rho_{total}$ 
 and the inhomogeneity $\delta r_\varphi \equiv  \delta \rho_\varphi
/\rho_{total}$.
Note that initially ($t=0$) the total energy density is homogeneous
 ($\delta \rho_{total}(0)\equiv0$).
We are considering isocurvature perturbations at $t=0$. 
The perturbation caused by the fluctuation of the field
$\varphi$ ($\delta \rho_{\varphi}/\rho_{total}
\sim r_\varphi\times (\delta r_\varphi/r_\varphi)$) 
may be suppressed when $r_\varphi \ll 1$.
However, if $\rho_\varphi$ scales milder than other components of
the Universe, the ratio $r_\varphi$ may grow and may finally reach
$r_\varphi\sim 1$ after a time period.  
The typical example is curvatons, whose ratio grows after inflation
and finally dominates the density of the Universe before decay.
Usually the curvaton scales as a matter during curvaton oscillation,
but in some extended models the curvatons can be associated with
topological defects\cite{Topolo-curv} or hybrid
states\cite{Hybrid-Curvatons} 
 that do not scale as a matter.
In this section we discuss whether the warm scenario
can accommodate the usual curvaton scenario.

Assume that the curvaton $\varphi$ is identified with the warm-flat direction,
then the amplitude of the fluctuation $\delta \varphi$ will be enhanced if
the warm background is significant ($r_T\gg 1$).
This causes enhancement of the perturbations, as in the case of the
Affleck-Dine field.
If the dissipation is strong for the curvatons, the slow-roll period
will be
extended because of the large $r$ in the field equation.
This enhances the ``initial'' ratio of the curvaton density at the
beginning of the curvaton oscillation, since the energy density of the
curvaton is a constant before oscillation.
These effects enhance both $r_\varphi$ and $\delta r_\varphi$ 
at the beginning of the oscillation helping curvatons generate the
curvature perturbations. 
However, in contrast to the situation before the oscillation,
the evolution during the curvaton oscillation may cause a serious problem.
For the strongly dissipating curvatons
we have to consider a field $\varphi$ with interactions
between mediating field, where the dissipation is significant only when
the mediating field decays into light matter.
Since the typical interaction required for the strong dissipation is
nothing but the interaction required for instant preheating followed by
instant reheating\cite{IP-orig}, the first 
oscillation will lead to efficient production of the mediating field 
near the origin and subsequently the instant decay.\footnote{We will
discuss this issue later in sec.2.2.3 for the inhomogeneous preheating
scenario.} 
As a result, strongly dissipating curvatons cannot oscillate for a long
time.
The curvaton density cannot grow during the oscillation.
Since the oscillation of the dissipative curvaton cannot last long,
the ratio $r_\varphi$ associated with the warm curvaton cannot 
grow during oscillation.\footnote{
Heavy curvaton accompanied by phase
transition\cite{matsuda_curvaton} may avoid this problem.
Efficient production of stable Q-balls\cite{Topolo-curv} may also
avoid this problem\cite{MSSM-curvatons}.
These exceptional cases are not excluded but should be discussed
elsewhere. }
To understand the situation, we consider an optimistic case in which
strong dissipation ($r\gg 1$) delays the curvaton oscillation until
$H_{osc}\simeq r^{-1} m_\varphi$.
The ratio at the beginning of the oscillation is given by
\begin{equation}
\label{eq1}
r_\varphi \simeq \frac{m_\varphi^2 \varphi^2}{H_{osc}^2 M_p^2}
\simeq  r^2\frac{\varphi^2}{ M_p^2},
\end{equation}
which suggests that $r_\varphi\sim 1$ can be reached before the
oscillation.
The curvaton decays before cosmological scales start to enter the
horizon.
The curvature perturbation is given by
\begin{equation}
\zeta = r_\varphi\zeta_\sigma\equiv r_\varphi
\frac{1}{3}\frac{\delta \rho_{\varphi}}{\rho_{\varphi}}. 
\end{equation} 
The spectrum of the curvature perturbation is given by
\begin{equation}
{\cal P}_\zeta^{1/2}=2r_\varphi \frac{\delta\varphi_I}{\varphi_I},
\end{equation}
where the subscript $I$ means that quantities are to be estimated
during inflation.
In the above equation we assumed that the curvaton potential is
quadratic: $V(\sigma)\simeq m_\varphi^2\varphi^2$.  
In order to generate the curvature perturbations of the Universe, the
curvaton fluctuation must (at least) satisfy the bound given by
\begin{equation}
\label{eq2}
r_\varphi\frac{\delta \varphi_I}{\varphi_I}\simeq r_\varphi
\left[r^{1/4}r_T^{1/2} \frac{H}{\varphi}\right]_I \sim 10^{-5},
\end{equation}
where $T$ represents the temperature when the
fluctuation exits horizon.
Combining Eq.(\ref{eq1}) and (\ref{eq2}), and assuming that $\varphi$
is constant until oscillation begins, we find
\begin{equation}
\varphi\sim M_p \frac{M_p\times 10^{-5}}{H_I}r^{-9/4}r_T^{-1/2}
\end{equation}
which suggests that warm curvatons may accommodate MSSM flat directions
even if it decays fast at the first oscillation.
The result is in contrast with the usual MSSM-curvaton model\cite{MSSM-curvatons}. 

\subsubsection{Inhomogeneous phase transition} 
In this section we consider an inhomogeneous phase transition.
The $i$-th cosmological period (we call this period a ``phase'' of the
Universe)   can be
characterized by the  
scaling property of the dominant component of the 
energy density $\rho_{dom}\propto a^{-n_i}$.
At the boundary of the ``phases'' in the time-direction, a gap
$\Delta n_i\equiv n_{i+1}-n_i$ appears.
The local delay of the ``phase transition'' is a natural source of
density perturbations\cite{IH-pt}.
There are at least three distinctive mechanisms of this kind:
\begin{enumerate} 
\item ``Generation of curvature perturbations at the end of
      inflation\cite{End-Modulated, End-multi, End-multi-mat0}''
 occurs for a change in $n_i$ from $n_i=0$ to $n_{i+1}=3$
 (inflaton decays into matter) or to $n_{i+1}=4$ (inflaton
      immediately decays into radiation).
\item ``Inhomogeneous reheating'' occurs for a change in $n_i$ from 
 $n_i=3$ to $n_{i+1}=4$ (matter decays into radiation).
\item ``Inhomogeneous phase transition'' is possible for the
      conventional phase transition in the early Universe\cite{IH-pt}.
\end{enumerate}
The phase boundary is inhomogeneous if the field value ($\varphi$) that
determines the boundary is inhomogeneous 
($\delta \varphi \ne 0$).
For instance, consider the potential for hybrid inflation given by
\begin{equation}
V=V_0 +\frac{1}{2}m_\phi^2\phi^2 -\frac{1}{2}m_\chi^2\chi^2
+\frac{1}{4}\lambda \chi^4+\frac{1}{2}\lambda_\phi\phi^2\chi^2
+\frac{1}{2}\lambda_\varphi\varphi^2\chi^2+V_2(\varphi),
\end{equation}
where $\phi$ and $\chi$ are the inflaton and the waterfall field.
This is the same as the original hybrid model except for the
additional light field $\varphi$.
Here $\varphi$ is distinguished from the inflaton field.
Assuming that $\varphi$ is light during inflation, inflation ends when
$\chi$ is destabilized at $\phi=\phi_e$, which corresponds to
\begin{equation}
\lambda_\phi\phi_e^2+\lambda_\varphi\varphi^2=m_\chi^2.
\end{equation}
In this case the fluctuation $\delta \varphi$ associated with 
the additional light field $\varphi$
causes inhomogeneous end of hybrid inflation and leads to 
$\delta N\ne 0$.
This is the basic mechanisms for generating curvature perturbations at
the end of inflation.
The mechanisms 1. 2. 3. will work with warm scenario 
in which $\varphi$ (the field that sources cosmological perturbation)
is identified with the warm direction.
$\varphi$ may be a moduli in the string theory.

Compared with the usual (non-dissipative) scenario, 
the most significant discrepancy will appear in the slow-roll condition
and the amplitude of the field fluctuations.
Namely, if the dissipation is strong for the field $(r\gg 1)$, the
$O(H^2)$ mass-term correction from the supergravity interactions does
not cause the $\eta$ problem for the field $\varphi$. 
This is the first significant advantage of the warm scenario.
Moreover, if the warm background is significant for the field, the
amplitude of the field fluctuations ($\delta \varphi$) will be enhanced.
This is the second advantage of the warm scenario.
As a consequence, the curvature perturbations caused by the warm
direction may depend on the dissipative coefficient $\Upsilon$
(i.e. it depends not only on the scalar potential but also 
on the interactions) and the temperature.

As far as one is considering the usual non-dissipative scenario,
directions that are not protected from the $O(H^2)$ supergravity
corrections are not considered as the source of the 
cosmological perturbations.
We claim that what is important in solving the $\eta$ problem is 
not the conventional mass protection but the interactions
 of the field that
 cause dissipative motion.
In fact, there can be many flat directions
that are not protected from $O(H)$ corrections but can roll slowly due
to the strong dissipation caused by the interactions.
Moreover, considering warm background during inflation, the amplitudes
of the field fluctuations may not be uniform but characterized by their
interactions. 
Therefore, the cosmological perturbations sourced by these warm
directions are potentially very important for analyzing the interactions
of the underlying theory.
In contrast to the usual non-dissipative scenario, the cosmological
perturbations in the warm scenario depend not only on the potential but
also on the interactions of the field.

\subsubsection{Inhomogeneous number density from preheating}
Preheating uses typical interaction given by
\begin{equation}
{\cal L}_{int}\simeq \frac{g_{PR}^2}{2} \phi^2 \varphi^2
\end{equation}
and oscillation of the field $\phi$.
Then the interaction leads to efficient production of
$\varphi$ near $\phi=0$.
Assuming so-called instant preheating scenario,
the number density of $\varphi$ created by preheating is given by
\begin{equation}
\label{n_chi}
n_\varphi = \frac{(g_{PR}|\dot{\phi}(t_*)|)^{3/2}}{8\pi^3}
\exp\left[-\frac{\pi m_\varphi^2}{g_{PR}|\dot{\phi}(t_*)|}
\right],
\end{equation}
where $\phi$ is the oscillating field and 
$t_*$ denotes the time when $\phi$ reaches its minimum at $\phi=0$.
The mass ($m_\varphi$) of the preheat field $\varphi$ is assumed to be
very small near the enhanced symmetric point(ESP) at $\phi=0$.
In order to create inhomogeneities in the number density of the
preheat field $\varphi$, the mass of the preheat field
must be inhomogeneous in space.
Typically, a light field $\phi_l$ is introduced in addition to the
oscillating field $\phi$, with the interaction $\sim g_{PR}^2
\varphi^2(\phi^2+\phi_l^2)$. 
Then, the mass of the preheat field is given by
\begin{equation}
m_\varphi^2 \simeq g_{PR}^2 (\phi^2+\phi_l^2),
\end{equation}
which is inhomogeneous for $\delta \phi_l \ne 0$ at the ESP($\phi=0$).

The light field $\phi_l$ can be identified with dissipative direction,
if it has interactions with 
mediating fields $\chi_i$ that decay into light particles.
The significant dissipation of the field $\phi_l$ does not cause serious
problem to the original inhomogeneous preheating model.
As a consequence, our warm-flat scenario accommodates inhomogeneous
preheating, solving the $\eta$-problem for the field $\phi_l$ 
and enhancing the amplitude of the perturbations $\delta \phi_l$
when inflation is warm.

In the above discussion we showed that dissipative direction $\phi_l$
accommodates inhomogeneous preheating scenario.
However, if the instant decay after preheating is assumed for the
$\phi$-field oscillation, which is the case in the first model of
inhomogeneous preheating proposed in Ref.\cite{preheating-application}, 
{\bf the dissipative effect may not be negligible for the motion of
the oscillating field $\phi$}. 
In fact, if the instant decay is assumed for the preheat field $\varphi$, 
$\varphi$ can play the role of the mediating field.
Then, significant dissipation is expected before preheating,
which typically leads to the delay of the $\phi$ oscillation. 
To show explicitly the dissipative effect that may be significant for
 $\phi$, we consider the potential with  mass hierarchy ($m_l\ll m$);
\begin{equation}
V(\phi, \phi_l)=\frac{1}{2}m_{l}^2\phi_l^2 +\frac{1}{2}m^2\phi^2,
\end{equation}
and the dissipation based on the zero-temperature approximation.
The interaction with the preheat field is given by
\begin{equation}
{\cal L}_{int}\simeq \frac{g_{PR}^2}{2} \phi^2 \varphi^2,
\end{equation}
where the preheat field $\varphi$ causes dissipation.
The decay into light fermions, which is needed for the preheating
scenario followed by the instant
decay, is induced by the term
\begin{equation}
{\cal L}_{\psi\varphi}=h \varphi\bar{\psi}\psi.
\end{equation}
Obviously, for $g\sim h\sim O(1)$, the dissipative process
$\phi\rightarrow \varphi \rightarrow \psi$ is efficient for the model.
Typically, the dissipative coefficient in the zero-temperature
approximation is given by
\begin{equation}
\Gamma \simeq 10^{-2}g_{PR}^3 h^2 m_\varphi.
\end{equation}
Then the effective slow-roll conditions for the field $\phi$ are given
by 
\begin{eqnarray}
\epsilon_w &\simeq& M_p^2 
\left(\frac{m^2 \phi}{V}\right)^2 
\frac{1}{(1+r)^2}
\sim  
\left(\frac{m}{10^{-2}\times  \phi }\frac{1}{g_{PR}^3 h^2}\right)^2 < 1
\\
\eta_w &\simeq& \frac{m^2}{H^2}
\frac{1}{(1+r)^2}
\sim \frac{m^2}{10^{-4}\times  \phi^2}\frac{1}{g_{PR}^6h^4}
<1,
\end{eqnarray}
which lead to simple condition given by $\phi> 10^{2} 
\times m/g_{PR}^{3}h^2$,
where the motion of the field $\phi$ is slow
because of the dissipation caused by the preheat field.
Therefore, the requirement for instant preheating followed by
instant reheating leads to a new condition for the model, which 
affects the initial condition of the oscillation.
Namely, the amplitude of the $\phi$-field oscillation is
not given by the usual condition $\phi\simeq M_p$, but by 
the new condition $\phi\simeq  Min[10^{2} \times m/g_{PR}^{3}h^2, M_p]$
that is obtained considering the dissipative effect caused by the
preheat field $\varphi$.

\section{Conclusions and discussions}
In this paper we considered cosmological scenarios associated with the
dissipative scalar fields that are not identified with the inflaton field.
The strong dissipation ($r>1$) solves the $\eta$-problem associated with
the supergravity interactions. 
Field fluctuations of the dissipative field may be enhanced
by the warm background during inflation.
We investigated the possibility that these effects may alter the usual
cosmological scenarios associated with SUSY-flat directions.  
For an obvious example we first considered the Affleck-Dine mechanism,
then briefly discussed several mechanisms of generating curvature
 perturbations in which the dissipative field sources the cosmological
 perturbations. 
Our point is that in generic situations there are many flat directions
that are not protected from supergravity corrections but can roll slowly
due to the strong dissipation caused by interactions with non-flat
directions. 
The dissipative field is not the inflaton field, but can source
cosmological perturbations.
Since the cosmological perturbations created by the dissipative field
are characterized by their interactions, they are potentially very
important for analyzing the interactions of the underlying theory
using cosmological observations.

\section{Acknowledgment}
We wish to thank K.Shima for encouragement, and our colleagues at
Tokyo University for their kind hospitality.
\appendix
\section{Calculation of the dissipative coefficient at zero temperature}
A key mechanism of dissipative motion, which is generic in realistic
flat direction involves the scalar flaton $\phi$ exciting a heavy
bosonic field (mediating field) $\chi$ which decays to light fermions
$\psi_d$.
This mechanism is expressed by the typical interaction
Lagrangian, which is given by
\begin{equation}
{\cal L}_{int}=-\frac{1}{2}g^2 \phi^2\chi^2 -h\chi\bar{\psi}\psi,
\end{equation}
where $\psi_d$ are light fermions to which the mediating field $\chi$
can decay (i.e. $m_\chi>2m_\psi$ at the decay.).
For typical MSSM directions, the mediating fields $\chi$ are the
heavy gauge bosons, Higgses or scalar supersymmetric partners that
obtain large mass from $\phi$.
The light fermions are neutrinos (which is massless in MSSM),
fermion supersymmetric partners of flat directions or any other
particles with the mass satisfying $m_\chi>2 m_\psi$.

A significant example\cite{gamma-MS} leading to a non-trivial flaton
dissipation  
is for $m_\chi> 2 m_\psi$ and $m_\chi > m_\phi$,
where $\chi\rightarrow 2\psi$ is possible but the direct decay
``$\phi\rightarrow$other'' is not allowed.
Nevertheless, it is not difficult to imagine the physics underlying the
dissipation of the field $\phi$.
It is useful to look into these processes involving $\phi$ that have an
imaginary term associated to dissipation.
The imaginary term is interpreted in terms of an effective theory for
$\phi$, which appears after integrating over other fields.
First integrating $\psi$ that only couples to $\chi$, the $\chi$
propagator is dressed, which has a real part (a shift in the mass) and
an imaginary part (rate of the decay).
Then integrating over the dressed $\chi$, the relevant contributions to
the dissipative mechanism appear from the decay of $\chi$.\footnote{The
original calculation given by Morikawa and Sasaki\cite{gamma-MS} is
different from the above description.
Their calculation is based on the Bogoliubov transformation with the
time-dependent background $\dot{\phi}\ne 0$ and the quartic interaction
$\propto \phi^4$. 
The mediating field is not considered in the original calculation.} 
For the decay rate
\begin{equation}
\Gamma_{\chi}\simeq \frac{N_\psi}{8\pi} h^2 m_\chi
\simeq \frac{N_\psi}{8\pi} h^2 g \phi,
\end{equation}
where $N_\psi$ is the number of the light fermions,
the dissipative coefficient is given by\cite{gamma-MS}
\begin{equation}
\Gamma \sim N_\chi 
\frac{\sqrt{2}g^3  N_\psi h^2\phi}{8^3\pi^2},
\end{equation}
where $N_\chi$ is the number of the mediatinge fields.\footnote{See
ref.\cite{gamma-MS} for details of the calculation.}

\end{document}